# Electric and Magnetic Excitation of Coherent Magnetic Plasmon Waves in a One-dimensional Meta-chain

C. Zhu[1], H. Liu[1,*], S. M. Wang[1], T. Li[1], J. X. Cao[1], Y. J. Zheng[1], L. Li[1], Y. Wang[1], S. N. Zhu[1] and X. Zhang[2]

[1]*Department of Physics, National Laboratory of Solid State Microstructures, Nanjing University, Nanjing 210093, China*

*liuhui@nju.edu.cn*

[2]*5130 Etcheverry Hall, Nanoscale Science and Engineering Center, University of California, Berkeley, California 94720-1740, USA*

**Abstract:** A one-dimensional diatomic meta-chain with equal-size holes and different-length slits is designed. Broadband coherent magnetic plasmon waves (MPW) are formed in such a system, excited by both the electric resonance in the slits and the magnetic resonance in the holes in a wide range of incidence angles ($0^0$–$40^0$) and broad frequency bands (200–230 THz). The dispersion properties of the MPW measured in our experiments agree with the theoretical calculation based on the Lagrange model. The coherent MPWs reported in this paper may have applications in subwavelength integrated nanocircuits.

## 1.Introduction

Although the invention of metamaterials has stimulated the interest of many researchers and has important applications in negative refraction [1–6], invisible cloaking [7–9], and many other transform optical designs, the basic design idea is very simple: composing effective media from many small structured elements and controlling their artificial electromagnetic (EM) properties. According to the effective media model, the coupling interactions between the elements in metamaterials are somewhat ignored; therefore, the effective properties of metamaterials can be viewed as the "averaged effect" of the resonance property of the individual elements. However, the coupling interaction between elements should always exist when they are arranged into metamaterials. Occasionally, especially when the elements are very close, this coupling effect is not negligible and will have a substantial effect on the properties of the metamaterials. Recent studies have shown that the resonance coupling effect between split-ring resonators (SRR) can introduce magnetoinductive waves[10-16], electroinductive waves[17, 18], planar transmission lines[19], stereometamaterials[20-22] and radiation suppression[23]. Besides SRRs, the near-field coupling between other structures, such as nano-rods[24], fish-net[25] and nanosanwiches[26] are also reported.

In our previous theory work [27], another kind of coupling mechanism between resonance elements,

called exchanging current interaction, was proposed in a one-dimensional chain of SRRs. Compared with the near-field magnetic and electric coupling, this interaction is much stronger and can lead to coherent magnetic plasmon waves with broad dispersive frequency bands and a slow group velocity of light. In our experimental work [28], a diatomic chain of SHRs was devised with a unit cell, including two SHRs with equal-length slits and different-sized holes. The extraordinary transmission peaks induced by the excitation of the coherent MPWs in such a system can be directly observed in our experiment. However, these MPWs can only be excited through magnetic resonance in the nanoholes. Electric resonance does not contribute to the excitation. The normal incidence wave cannot be coupled onto MPWs, and the incidence angle has to be oblique.

In this letter, we propose a new design for the meta-chain of SHRs. Here, the unit cell includes two SHRs with different-length slits and equal-sized holes, which is different from our former work. The advantage of this new design is that the coherent magnetic plasmon wave can be excited both by the magnetic resonance in holes and the electric resonance in slits. Due to the strong electric resonance in the slits, the coherent magnetic plasmon in meta-chain can be excited much more efficiently. The excitation can also be realized under normal incidence. The incidence excitation angle can then be tuned in a wide range, from normal incidence to $40^0$, and a continuous wide excitation frequency band can be obtained through tuning the incidence angle. The measured dispersion of the coherent magnetic plasmon waves agrees with our theoretical calculated results.

## 2. Numerical Model and experimental results

Fig. 1(a) shows the structure of our designed slit-hole resonator, which is comprised of a nanohole and a nanoslit connecting the hole and the edge. The bulk material is silver. Compared with Pendry's SRR structure and other magnetic resonance structures, the SHR structure is much easier to fabricate with the focus ion beam (FIB) technique, and its resonance frequency is easily realized in the infrared range. This structure can be described by an equivalent LC circuit method, as shown as Fig. 1(b). The slit in the SHR can be seen as a capacitor, and the nanohole can be seen as a conductor that connects it. The Lagrangian of SHR structure can be written as $\Im = L\dot{Q}^2/2 - Q^2/(2C)$, where L is the inductance of the hole, Q is the total oscillation charge in the SHR, and C is the capacitance of the slit.

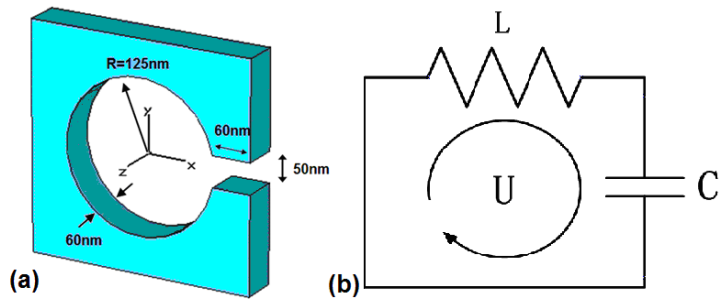

Fig. 1. Structure (a) and equivalent LC circuit (b) of a single SHR.

Based on this SHR resonator, a diatomic chain is designed, as shown in Fig. 2(a). The unit cell of this chain is composed of two SHRs with different-length slits and equal-sized holes. Here, the radius of the two holes are both 125 nm and the lengths of the two slits are 60 and 180 nm. The bulk material is silver, with a thickness of 60 nm. A sample of one-dimensional chain of SHRs was fabricated with the FIB technique. A sample is shown in Fig, 2(c). In our optical measurement, the sample is set on a rotation table with a y-polarized (E-field along y direction) light. The transmission is collected by an optical spectrum analyzer (ANDO, AQ-6315A) via a fiber coupler. Based on the method reported by Decker [15], the transmission spectra were taken by changing the incidence angle with this setup. The results are shown in Fig. 3(a). In this figure, we can see that the extraordinary optical transmission (EOT) peak is induced by MPW, changing from 230 to 200 THz, by tuning the incidence angle $\theta$ from $0^0$ to $40^0$. In this measurement, EOT can be obtained for the normal incidence wave ($\theta=0^0$), which did not occur in our previous work [28]. In the following model, we show that this is caused by the electric resonance in the slits.

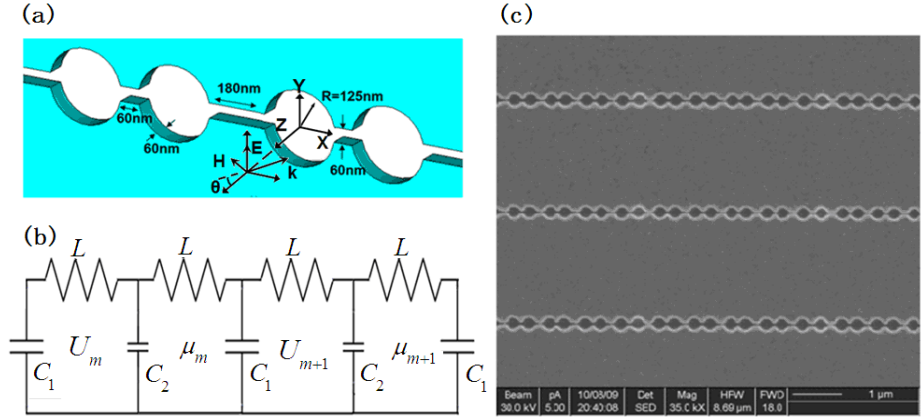

Fig. 2. Structure (a) and equivalent LC circuit of a SHR meta-chain;

(c) FIB image of the fabricated sample.

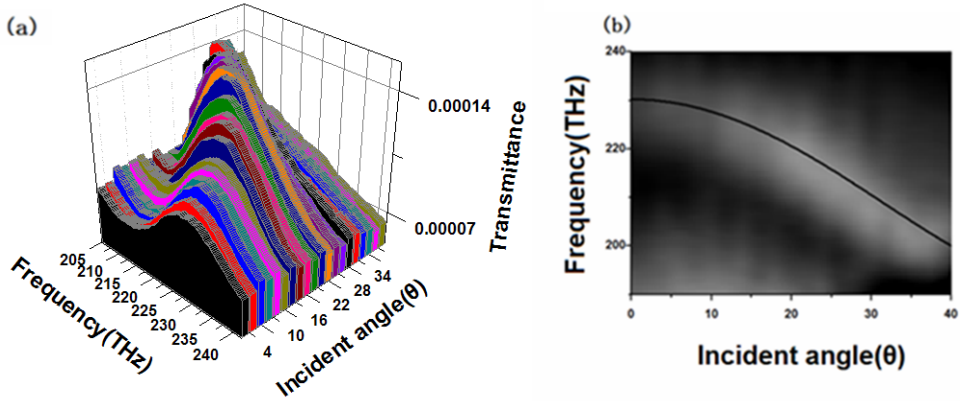

Fig. 3. (a).Observed transmission spectra under different incident angles; (b).The Measured transmission map and the calculated angular dependence curve of the optical mode of MPW.

## 3. Simulation results and discussions

The resonance frequency of the two SHRs can also be simulated numerically using the finite-difference time-domain method (FDTD) with a commercial software package (CST Microwave Studio). The longer SHR with a 180 nm slit has a resonance frequency of $\omega_1 = 135 THz$. Using the same method, the resonance frequency of the SHR with a 60 nm slit can also be found at $\omega_2 = 91 THz$. For the infinite diatomic SHR chain, its equivalent LC circuit could be described as Fig. 2(b), in which the longer SHR has a capacitance of $C_1$ and the shorter SHR has a capacitance of $C_2$. Therefore, the Lagrangian equation of the chain could be expressed as

$$\Im = \sum_m \left( \frac{L\dot{Q}_m^2}{2} + \frac{L\dot{q}_m^2}{2} - \frac{(Q_m - q_{m-1})^2}{2C_1} - \frac{(Q_m - q_m)^2}{2C_2} \right) \quad (1)$$

Here, we define $Q_m$ as the oscillation charge of the m-th longer SHR and $q_m$ as the oscillation charge of the m-th shorter SHR. Based on the Euler-Lagrangian equation, the dispersion of MPW could be obtained as

$$\omega_\pm^2 = (\omega_1^2 + \omega_2^2) \pm \sqrt{(\omega_1^4 + \omega_2^4) + 2\omega_1^2 \omega_2^2 \cos(kd)} \quad (2)$$

where k is the wave vector, $\omega$ is the angular frequency, and d=740 nm is the period of the chain. In the above simulations, we have already obtained $\omega_1 = 135 THz$, $\omega_2 = 91 THz$. Substituting them into Eq. (2) we can obtain two different dispersion branches, which are depicted as two separate black curves in Fig. 4(a). CST simulations show that the two branches stand for two different MPW modes by normal incidence. For the upper branch $\omega_+$, the neighboring SHR units oscillate in the opposite phase [see Fig. 4(b)], which can be called the optical branch, as reported by Sydoruk [29]. For the lower branch $\omega_-$, the neighboring units oscillate in the same phase [see Fig. 4(c)], which can be called the acoustic branch. The light in the free space is depicted as the blue line in Fig. 4(a). The optical branch of MPW is above the blue line; therefore, it can be excited by the plane wave from the far-field with an oblique incidence angle. This introduces an extraordinary optical transmission in the experiment. To compare our experimental results with the Lagrange theory, the dispersion of MPW under different incidence angles was calculated from Equation (2) based on the wave vector matching condition $k = k_0 \sin\theta$. The theoretical results are shown as the black solid curve in Fig. 3 (b), which agree with our measurements.

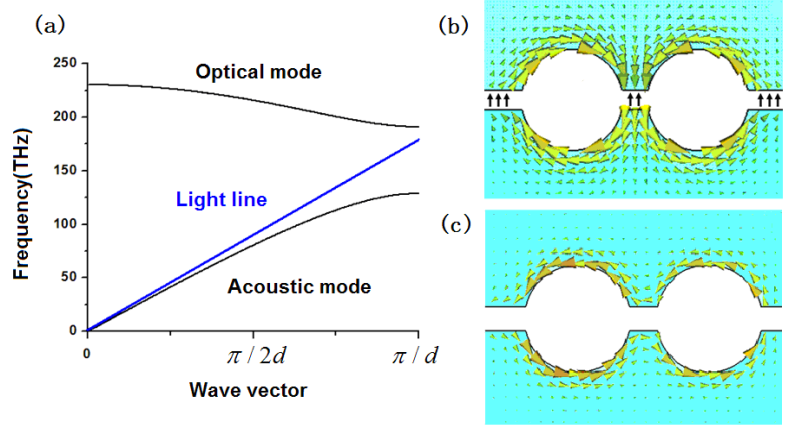

Fig. 4. (a). Dispersion properties of MPW; (b). Current distribution of induced current in one unit cell for the optical mode, in which the arrays represent the direction and the intensity of the incident electric field; (c). Current distribution for the acoustic mode.

In our former work, the SHR chain had equal-length slits and different-sized holes, and its EOT can only be excited for the incidence angle $\theta \geq 6^0$. In this work, for the normal incidence wave ($\theta = 0^0$), an EOT resonance peak is observed at 230 THz [see Fig. 3(a)]. This is caused by the electric resonance in the SHR, which was not reported in our former work. When the light is normally incident on the sample, no magnetic field penetrates the hole; therefore, the magnetic resonance does not contribute to the MPW. Only the electric field, which is always in the y direction, excites an electric resonance in the slits. If the neighboring slits are of equal lengths, as in our former work, the electric resonances in the two slits will have equal strengths and cancel out each other. Then, the electric resonance cannot contribute to the excitation of MPW for the normal incidence wave. For the design structure, the two neighboring slits have different lengths; therefore, the electric resonances have different strengths and do not cancel out each other [see Fig. 4 (b)]. The final contribution of electric resonance to MPW should be that the electric resonance of the longer slit subtracts that of the shorter slit. When the incidence angle increases, increasing magnetic flux will pass through the hole, and the magnetic resonance in the holes will become stronger. As a result, the EOT peak becomes stronger when the incidence angle increases from $0^0$ to $24^0$, as shown in Fig. 3(a). For $\theta \geq 24°$, the EOT peak will begin to decrease because the projected area of the fabricated pattern in the incidence direction decreases and the whole transmitted energy is reduced.

In this work, the experimental results show that the optical branch of MPW can be excited both by electric and magnetic resonance from far-field incidence waves. However, the acoustic MPW mode cannot be coupled to the far-field because its dispersion curve is below the light line [see Fig. 4 (a)]. Some near-field techniques have to be used to excite this mode, such as scanning near-field optical microscopy (SNOM). However, without coupling with far-field waves, the radiation loss of this mode could be very low, making it useful for subwavelength energy transport. Further experimental investigation on the SHR chain should be conducted in the future.

## 4. Conclusion

In summary, we have proposed and studied the coherent magnetic plasmon mode in a one-dimensional meta-chain. Our theoretical calculations and experimental results have proven that the coherent MPW in such a structure can be excited not only by the magnetic resonance, but also by the electric resonance; therefore, it has better properties compared with our former structure. The resonance frequency of the excited MPW can be easily tuned in a broad width by directly changing the incidence angle. A Lagrange model was employed to describe the coherent mode, and the calculated results agree with the experimental results. The coherent MPWs reported in this paper may have applications in subwavelength integrated nanocircuits.

**Acknowledgements**

This work is supported by the National Natural Science Foundation of China (No.10704036, No.10874081, No.60907009, No.10904012, No.10974090 and No. 60990320), and by the National Key Projects for Basic Researches of China (No. 2006CB921804, No. 2009CB930501 and No. 2010CB630703).

**References and links**

1. R. A. Shelby, D. R. Smith, and S. Schultz, "Experimental Verification of a Negative Index of Refraction," Science 292, 77-79 (2001).
2. S. Linden, C. Enkrich, M. Wegener, J. Zhou, T. Koschny, and C. M. Soukoulis, "Magnetic Response of Metamaterials at 100 Terahertz," Science 306, 1351-1353 (2004).
3. S. Zhang, W. Fan, N. C. Panoiu, K. J. Malloy, R. M. Osgood, and S. R. J. Brueck, "Experimental Demonstration of Near-Infrared Negative-Index Metamaterials," Physical review letters 95, 137404 (2005).
4. J. Valentine, S. Zhang, T. Zentgraf, E. Ulin-Avila, D. A. Genov, G. Bartal, and X. Zhang, "Three-dimensional optical

metamaterial with a negative refractive index," Nature 455, 376-379 (2008).

5. V. M. Shalaev, W. Cai, U. K. Chettiar, H.-K. Yuan, A. K. Sarychev, V. P. Drachev, and A. V. Kildishev, "Negative index of refraction in optical metamaterials," Opt. Lett. 30, 3356-3358 (2005).

6. G. Dolling, C. Enkrich, M. Wegener, C. M. Soukoulis, and S. Linden, "Simultaneous Negative Phase and Group Velocity of Light in a Metamaterial," Science 312, 892-894 (2006).

7. J. B. Pendry, D. Schurig, and D. R. Smith, "Controlling Electromagnetic Fields," Science 312, 1780-1782 (2006).

8. R. Liu, C. Ji, J. J. Mock, J. Y. Chin, T. J. Cui, and D. R. Smith, "Broadband Ground-Plane Cloak," Science 323, 366-369 (2009).

9. J. Valentine, J. Li, T. Zentgraf, G. Bartal, and X. Zhang, "An optical cloak made of dielectrics," Nat Mater 8, 568-571 (2009).

10. E. Shamonina, V. A. Kalinin, K. H. Ringhofer, and L. Solymar, "Magnetoinductive waves in one, two, and three dimensions," Journal of Applied Physics 92, 6252-6261 (2002).

11. O. Sydoruk, A. Radkovskaya, O. Zhuromskyy, E. Shamonina, M. Shamonin, C. J. Stevens, G. Faulkner, D. J. Edwards, and L. Solymar, "Tailoring the near-field guiding properties of magnetic metamaterials with two resonant elements per unit cell," Physical Review B 73, 224406 (2006).

12. A. Radkovskaya, O. Sydoruk, M. Shamonin, E. Shamonina, C. J. Stevens, G. Faulkner, D. J. Edwards, and L. Solymar, "Experimental study of a bi-periodic magnetoinductive waveguide: comparison with theory," IET Microwaves, Antennas & Propagation 1, 80-83 (2007).

13. I. V. Shadrivov, A. N. Reznik, and Y. S. Kivshar, "Magnetoinductive waves in arrays of split-ring resonators," Physica B: Condensed Matter 394, 180-183 (2007).

14. N. Liu, and H. Giessen, "Three-dimensional optical metamaterials asmodel systems for longitudinal and transversemagnetic coupling," Opt. Express 16, 21233-21238 (2008).

15. M. Decker, S. Burger, S. Linden, and M. Wegener, "Magnetization waves in split-ring-resonator arrays: Evidence for retardation effects," Physical Review B 80, 193102 (2009).

16. M. Decker, S. Linden, and M. Wegener, "Coupling effects in low-symmetry planar split-ring resonator arrays," Opt. Lett. 34, 1579-1581 (2009).

17. M. Beruete, F. Falcone, M. J. Freire, R. Marques, and J. D. Baena, "Electroinductive waves in chains of complementary metamaterial elements," Applied Physics Letters 88, 083503-083503 (2006).

18. N. Liu, S. Kaiser, and H. Giessen, "Magnetoinductive and Electroinductive Coupling in Plasmonic Metamaterial Molecules," Advanced Materials 20, 4521-4525 (2008).

19. J. D. Baena, J. Bonache, F. Martin, R. M. Sillero, F. Falcone, T. Lopetegi, M. A. G. Laso, J. Garcia-Garcia, I. Gil, M. F. Portillo, and M. Sorolla, "Equivalent-circuit models for split-ring resonators and complementary split-ring resonators coupled to planar transmission lines," Microwave Theory and Techniques, IEEE Transactions on 53, 1451-1461 (2005).

20. N. Liu, H. Liu, S. Zhu, and H. Giessen, "Stereometamaterials," Nat Photon 3, 157-162 (2009).

21. H. Liu, D. A. Genov, D. M. Wu, Y. M. Liu, Z. W. Liu, C. Sun, S. N. Zhu, and X. Zhang, "Magnetic plasmon hybridization and optical activity at optical frequencies in metallic nanostructures," Physical Review B 76, 073101 (2007).

22. H. Liu, J. X. Cao, S. N. Zhu, N. Liu, R. Ameling, and H. Giessen, "Lagrange model for the chiral optical properties of stereometamaterials," Physical Review B 81, 241403.

23. T. Q. Li, H. Liu, T. Li, S. M. Wang, J. X. Cao, Z. H. Zhu, Z. G. Dong, S. N. Zhu, and X. Zhang, "Suppression of radiation loss by hybridization effect in two coupled split-ring resonators," Physical Review B 80, 115113 (2009).

24. J. Cao, H. Liu, T. Li, S. Wang, T. Li, S. Zhu, and X. Zhang, "Steering polarization of infrared light through hybridization effect in a tri-rod structure," J. Opt. Soc. Am. B 26, B96-B101 (2009).

25. T. Li, H. Liu, F. M. Wang, Z. G. Dong, S. N. Zhu, and X. Zhang, "Coupling effect of magnetic polariton in perforated metal/dielectric layered metamaterials and its influence on negative refraction transmission," Opt. Express 14, 11155-11163 (2006).
26. S. M. Wang, T. Li, H. Liu, F. M. Wang, S. N. Zhu, and X. Zhang, "Magnetic plasmon modes in periodic chains of nanosandwiches," Opt. Express 16, 3560-3565 (2008).
27. H. Liu, D. A. Genov, D. M. Wu, Y. M. Liu, J. M. Steele, C. Sun, S. N. Zhu, and X. Zhang, "Magnetic Plasmon Propagation Along a Chain of Connected Subwavelength Resonators at Infrared Frequencies," Physical review letters 97, 243902 (2006).
28. H. Liu, T. Li, Q. J. Wang, Z. H. Zhu, S. M. Wang, J. Q. Li, S. N. Zhu, Y. Y. Zhu, and X. Zhang, "Extraordinary optical transmission induced by excitation of a magnetic plasmon propagation mode in a diatomic chain of slit-hole resonators," Physical Review B 79, 024304 (2009).
29. O. Sydoruk, O. Zhuromskyy, E. Shamonina, and L. Solymar, "Phonon-like dispersion curves of magnetoinductive waves," Applied Physics Letters 87, 072501-072503 (2005).